\documentclass[twocolumn,twoside,slac_two]{revtex4}
\usepackage{graphicx}
\usepackage{fancyhdr}
\begin{document}
\voffset=-1.3cm

\title{Discovery and Identification of Extra Neutral Gauge Bosons at the LHC}

%

\author{Ross Diener$^1$, Stephen Godfrey$^{1,2}$ and Travis A. W. Martin$^1$}
\affiliation{$^1$Ottawa-Carleton Institute for Physics, 
Department of Physics, Carleton University, Ottawa, Canada K1S 5B6 \\ 
$^2$TRIUMF, 4004 Wesbrook Mall, Vancouver, Canada V6T 2A3}

\begin{abstract}
We present the discovery prospects for a $Z'$ resonance and 
some expected results for $Z'$ diagnostic measurements at the LHC.
The  discovery reach is highly dependent on the energy and luminosities that
may be attained at the LHC, and a number of energy and integrated luminosity scenarios
are presented.
In addition, the use of third generation quark final states for distinguishing between 
models and for measuring a forward-backward asymmetry is explored.
\end{abstract}

\maketitle

\thispagestyle{fancy}

\section{Introduction\label{Introduction}}
There is a general consensus 
that the Standard Model (SM) is a low energy effective theory and that
some sort of new physics exists at higher energies than we can currently explore.
Many models have been developed to describe potential new physics and explain electroweak 
symmetry breaking (EWSB), among which are extended gauge sectors, including models with 
extra $U(1)$ 
factors~\cite{Hewett:1988xc,Langacker:2008yv,Rizzo:2006nw,Leike:1998wr,Cvetic:1995zs}.
Grand Unified Theories with $E_6$ breaking to 
$SU(5)\times U(1)_\chi \times U(1)_\psi$~\cite{Hewett:1988xc} are 
one example, while the Left-Right Symmetric model~\cite{Mohapatra:uf}, 
$SU(2)_L \times SU(2)_R \times U(1)$, is another.
Other approaches to describing EWSB include the Little Higgs model and 
variants~\cite{ArkaniHamed:2002qy,Schmaltz:2004de}, theories 
with extra dimensions~\cite{Randall:1999ee}, Supersymmetry, and Technicolor and Topcolor models.

New TeV scale $s$-channel structures are common to many of these models of new physics.
If an $s$-channel resonance were discovered, the immediate task would be to try to identify
the underlying theory.
In this report, we describe some approaches to distinguish between the different extra neutral
gauge bosons ($Z^{\prime}$) that appear in models of new physics.
We start with an update on discovery limits of $Z^{\prime}$'s at the LHC for the low centre-of-mass 
energy and low luminosity scenarios in the early years of the 
LHC program~\cite{Godfrey:1994qk,Erler:2009jh,Salvioni:2009mt}.
We then describe the use of top and bottom quark final states to distinguish between 
models~\cite{Godfrey:2008vf,Rizzo:1998ut,Barger:2006hm,Frederix:2007gi,Mohapatra:1992tc}, 
and give a preliminary account of more recent studies of 
forward-backward asymmetries ($A_{FB}$)~\cite{Langacker:1984dc,Petriello:2008zr} including
the use of top and bottom quarks.

The models analyzed in this contribution are the 
E6 set of models (E6 $\chi$, $\psi$, $\eta$)~\cite{Hewett:1988xc}, 
the Left-Right Symmetric (LR, $g_L/g_R = 1$)~\cite{Mohapatra:uf} and 
Alternate Left-Right Symmetric models (ALR, $g_L/g_R = 1$)~\cite{Ma:1986we}, 
the Ununified Model (UUM, $\tan\phi = 0.5$)~\cite{Georgi:1989xz}, 
the Sequential Standard model (SSM), 
a Topcolor Assisted Technicolor model (TC2, $\tan\theta=0.577$)~\cite{Harris:1999ya,Hill:1991at}, 
the Littlest Higgs model (LH, $\cot\theta_H=1$)~\cite{ArkaniHamed:2001nc,ArkaniHamed:2002qy}, 
the Simplest Little Higgs model (SLH)~\cite{Schmaltz:2004de,ArkaniHamed:2001nc},
the Anomaly Free Simple Little Higgs model (AFSLH)~\cite{Han:2005ru,Kong:2003vm}, 
a 331 model~\cite{Barreto:2007mt,Dias:2005xj}, 
and generic models with group structures of 
$SU(2)_L \times SU(2)_H$ ($\sin\phi = 0.9$)
\cite{Chivukula:1994mn,Chivukula:1995gu,Simmons:1996ws,Malkawi:1996fs,Muller:1996qs,He:1999vp} 
and $U(1)_L \times U(1)_H$ ($\cos\phi = 0.9$)
\cite{Hill:1994hp,Lane:1995gw,Lane:1996ua,Lane:1998qi,Popovic:1998vb}.

\section{Discovery Limits of Extra Neutral Gauge Bosons at the LHC\label{Discovery_Limits}}

\begin{figure*}[ht]
\centering
\includegraphics[width=85mm]{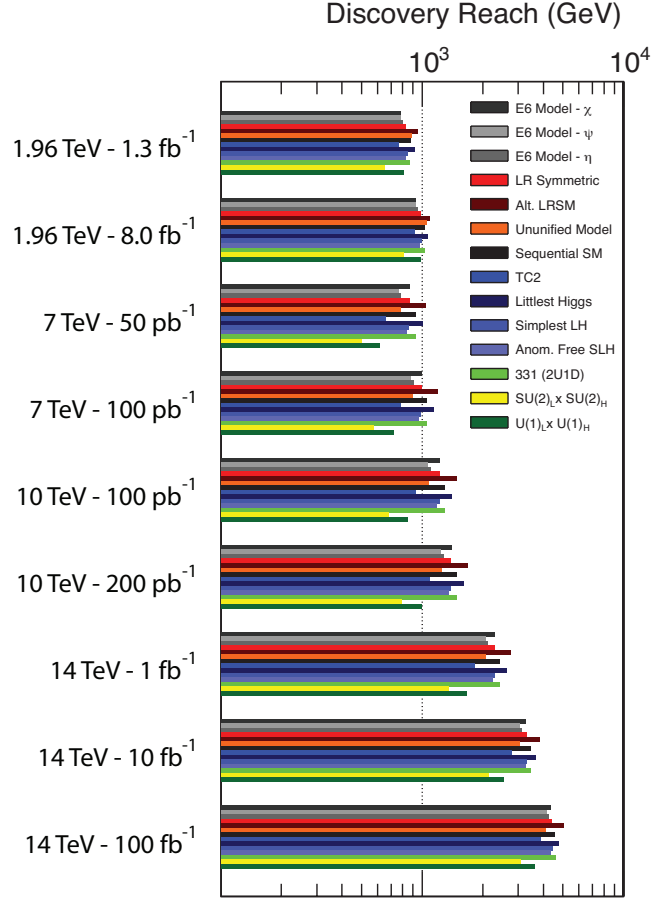}
\caption{Discovery reach for extra neutral gauge bosons at the LHC.} \label{fig:disc_reach}
\end{figure*}

An extra neutral gauge boson would be produced by a hadron collider via the Drell-Yan process 
and would show up as a resonant peak of events in the invariant mass distribution of the 
fermion-anti-fermion final state\cite{Godfrey:1994qk}.
For non-leptophobic $Z^{\prime}$'s, the cleanest search channel uses dilepton final states 
($\mu^+\mu^-$  or $e^+ e^-$)~\cite{Hewett:1988xc,Langacker:2008yv,Rizzo:2006nw,Leike:1998wr,Cvetic:1995zs}  
due to the small irreducible backgrounds and a clear, distinguishable
signal in the ATLAS~\cite{:2008zzm,:1999fq} and CMS~\cite{Ball:2007zza} detectors.
A small number of dilepton events clustered in an isolated invariant mass bin would be strong
evidence for the existence of an $s$-channel resonance.

To predict the discovery potential of the LHC for various models, we calculate the
cross section to dimuon final states for a given $Z^{\prime}$ mass and impose the following
kinematic cuts to reflect detector acceptances: $p_{T_l} > 20$~GeV and 
$|\eta_l|<2.5$~\cite{:2008zzm,:1999fq}.
In addition, we consider only contributions from events within an invariant mass bin around
the resonance mass peak, as defined by the ATLAS TDR~\cite{:1999fq}.
We define the discovery reach as the largest resonance mass, $M_{Z^\prime}$, that would
result in the observation of 5 events per dilepton channel, based on the luminosity provided.
The $Z^{\prime}$ discovery reach for various LHC energy and luminosity scenarios are shown in
Fig.~\ref{fig:disc_reach}.

Also included are estimates of the discovery reach for the same models at the Tevatron
assuming both 1.3~fb$^{-1}$ of integrated luminosity, as in Ref.~\cite{Aaltonen:2008vx}, 
and 8~fb$^{-1}$ to estimate the reach from the full expected luminosity.
In our analysis, we use similar detector acceptance and cuts as in Ref.~\cite{Aaltonen:2008vx}: 
we impose a kinematic cut of $p_{T} > 25$~GeV and consider events within two regions of
pseudorapidity - where both leptons satisfy $|\eta| < 1.1$, and where one lepton satisfies 
$|\eta| < 1.1$ and the other satisfies $1.2 < |\eta| < 2.0$.
As well, we consider only events within an invariant mass window of 
$|M_{Z^{\prime}}-M_{l^+l^-}| = \pm10\% M_{Z^{\prime}}$.  We use the discovery
criteria of 5 observed dilepton events as with the LHC study.
We note that our approach differs from that used by the CDF collaboration 
to obtain the current direct limits in Refs. \cite{Aaltonen:2008vx} and \cite{Aaltonen:2008ah}.

\section{$Z^{\prime}$ Identification Using $t$ and $b$ Quarks\label{Identification}}

The discovery of a TeV scale $s$-channel resonance would be a first step in a particle physics
revolution. 
But to determine the underlying physics will require the determination of the properties
of the resonance.
For dilepton final states, the observables that have been studied so far include:
the $Z^{\prime}$ cross section and width~\cite{Langacker:1984dc}, the angular distribution and 
centre-edge asymmetry~\cite{delAguila:1993ym,Dvergsnes:2004tw}, and
various analyses of the rapidity distributions through asymmetries and matrix 
analysis~\cite{delAguila:1993ym,Dittmar:2003ir,Petriello:2008zr,Li:2009xh}.

Quark final states were considered to have limited usefullness due to the inability to 
identify individual quark flavours in the final state.
However, $b$ and $t$ quarks can be uniquely identified~\cite{:2008zzm,:1999fq} 
from other jets in the final state, as can a small percentage of $c$ quarks.
We have shown that this ability to identify $t$ and $b$ quarks in the final state can be used
to discriminate between models.
The primary challenges of this analysis will be the identification efficiency and rejection
rates against reducible QCD backgrounds.

The recent ATLAS detector paper~\cite{:2008zzm} suggests $b$ quark tagging efficiencies of 
$\epsilon_b = 60\%$ are possible for high $p_T$ events at high luminosities, with a 100 to 1 
rejection against light and $c$ jets.
Requiring the tag for both the $b$ and $\bar{b}$ events reduces the efficiency, but improves
the rejection by another factor of 100 to 1.
Depending on the expected light jet backgrounds, the tradeoff between efficiency and rejection 
can be adjusted to improve signal significance - a higher rejection rate corresponds to a lower 
identification efficiency.

Top quarks decay rapidly to $b+W^+$, where the $W$ boson subsequently decays either to 
lepton ($e^+\nu_e$, $\mu^+\nu_\mu$, $\tau^+\nu_\tau$) or light quark ($u\bar{d}$, $c\bar{s}$)
final states.
The single lepton plus jets decay mode, 
$t\bar{t}\to W^+W^- b\bar{b} \to (l\nu)(jj)(b\bar{b})$, accounts for approximately $30\%$ of 
$t\bar{t}$ events, and is often considered to have the best signal to background ratio.
The CMS and ATLAS collaborations estimate an efficiency of 
$\epsilon_{t\bar{t}}\sim 2-5\%$~\cite{Khramov:2007ev,:2008zzm,:1999fq,D'hondt:2007aj},
but more recent studies suggest $\epsilon_{t\bar{t}}\sim 10\%$ may be possible.
A number of recent studies have considered purely hadronic modes, where both $W$ bosons
from the $t\bar{t}$ pair decay to light jets, and suggest rejection rates of $10^4$ may be
possible~\cite{Baur:2007ck,Kaplan:2008ie,Thaler:2008ju,Baur:2008uv}.
If these can be utilized, the total number of $t\bar{t}$ events that could be used would be
increased significantly.

\begin{figure}[htbp]
\centering
\includegraphics[width=65mm,clip]{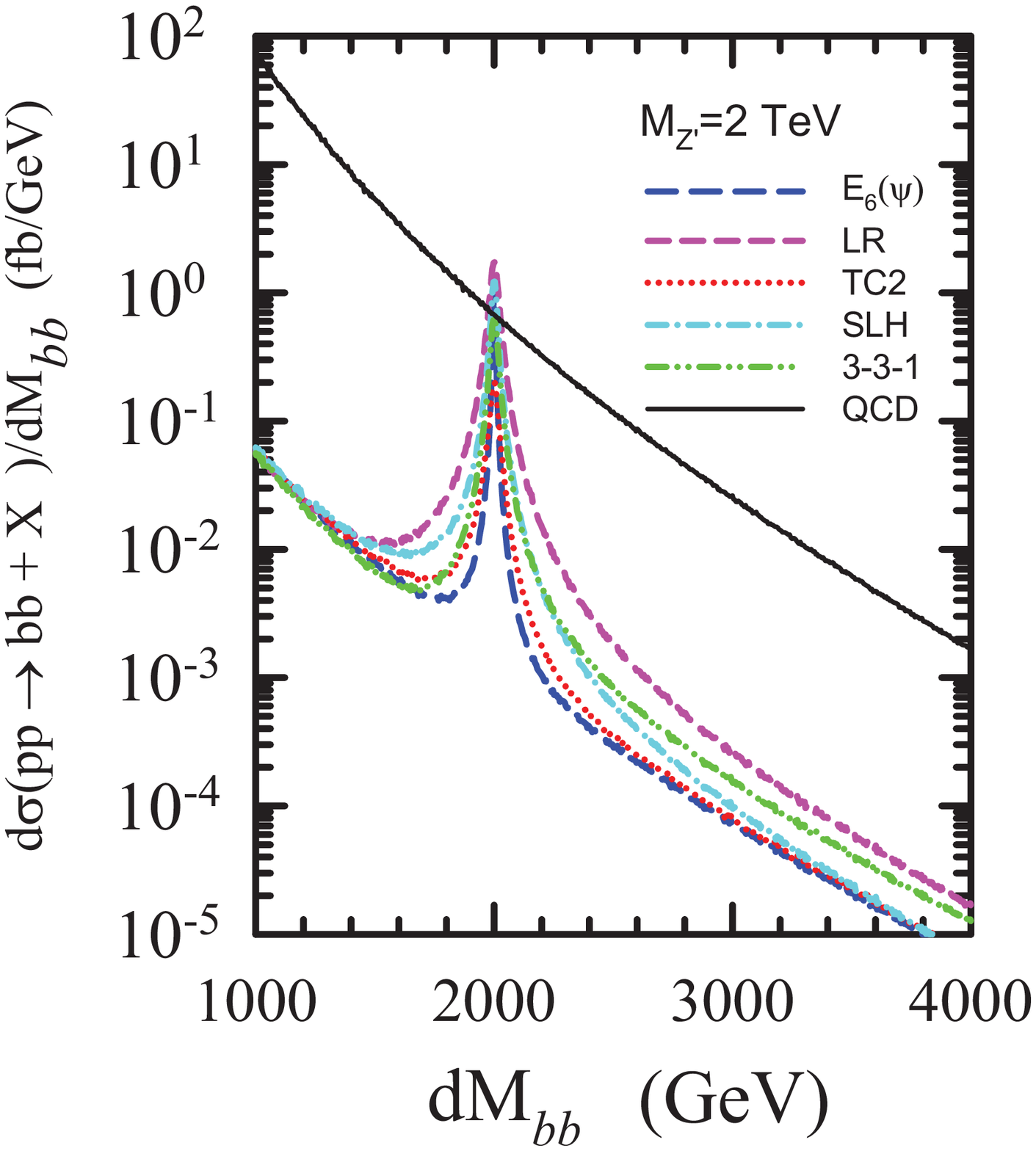} \\
\includegraphics[width=65mm,clip]{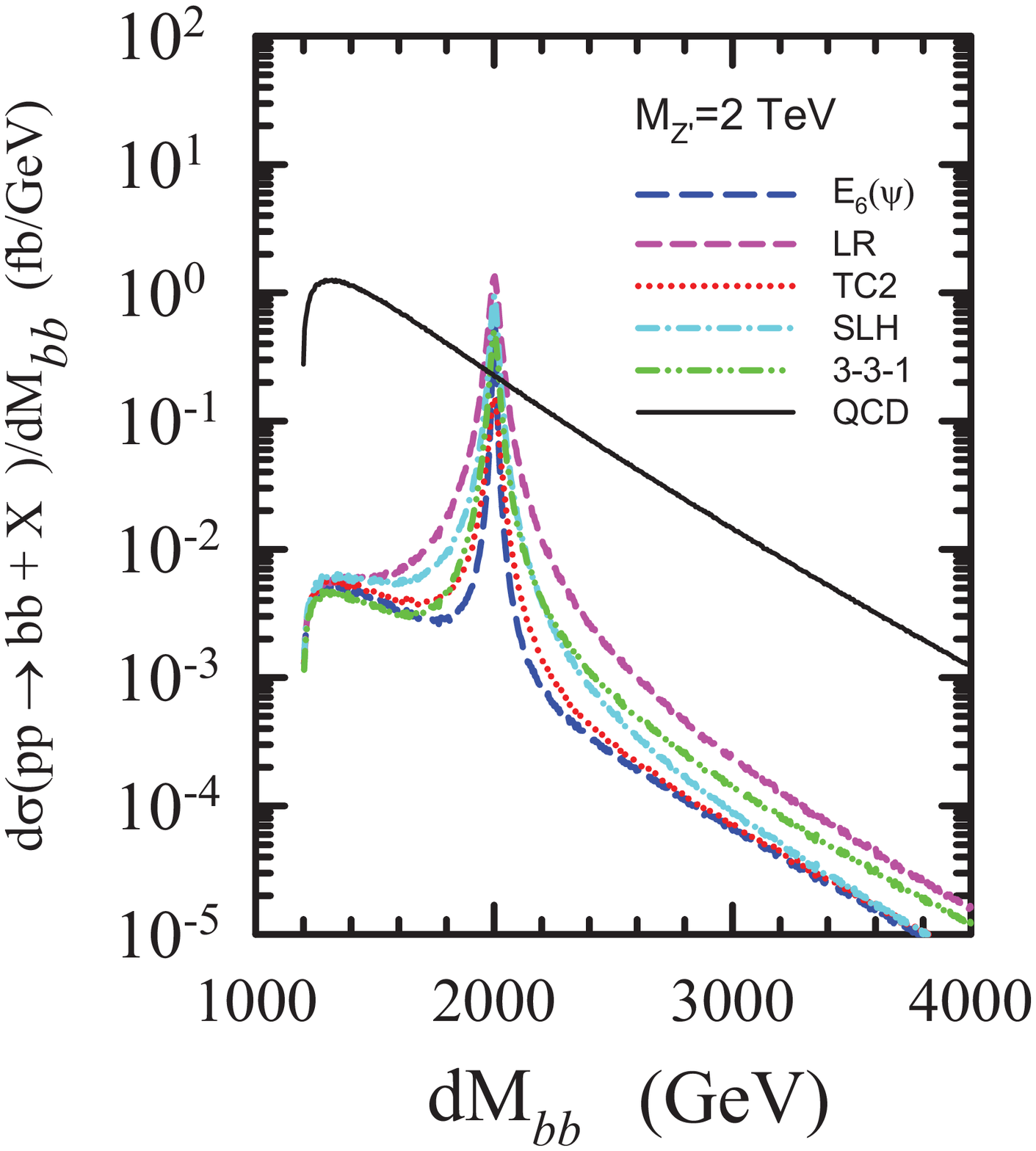}
\caption{Invariant mass distribution of $Z^{\prime}$ production at the LHC decaying to $b\bar{b}$ 
final states. Cuts include (a) $p_T~>~50$~GeV, $|\eta_l|~<~2.5$, and (b) $p_T~>~600$~GeV, 
$|\eta_l|~<~2.5$.\label{fig:im_bb}}
\end{figure}

Light jet rejection is not the only concern with measurements of electroweak processes involving
hadronic final states.
Even if large rejection rates are achieved, the irreducible SM QCD backgrounds are 
large~\cite{Baur:2007ck,Kaplan:2008ie,Thaler:2008ju,Baur:2008uv}.
Figure~\ref{fig:im_bb}(a) shows the the expected invariant mass distributions for 
a representative set of $Z^{\prime}$ models with $M_{Z^{\prime}} = 2$~TeV decaying to $b\bar{b}$ 
final states, 
including the expected SM backgrounds ($b\bar{b}$), without kinematic cuts.
We found that this background can be significantly reduced 
by imposing a cut on the $p_T$ of the reconstructed $t$ or $b$ jet~\cite{Godfrey:2008vf}.
We found that a value of $p_T > 0.3 M_{Z^{\prime}}$ provided a good balance between improving the
signal to background ratio while maintaining good total statistics.
The $b\bar{b}$ invariant mass distribution with this $p_T$ cut is shown in Fig.~\ref{fig:im_bb}(b),
where we can see an improvement in the visibility of the peak over the backgrounds.

Other important issues not considered in this study include non-QCD SM backgrounds, 
such as $Wb\bar{b}+jets$, $(Wb +W\bar{b})$, $W +jets$, and others. 
These backgrounds can be controlled by constraints on the cluster transverse mass and 
invariant mass of the jets.

\begin{figure*}[htbp]
\centering
\includegraphics[width=60mm,clip]{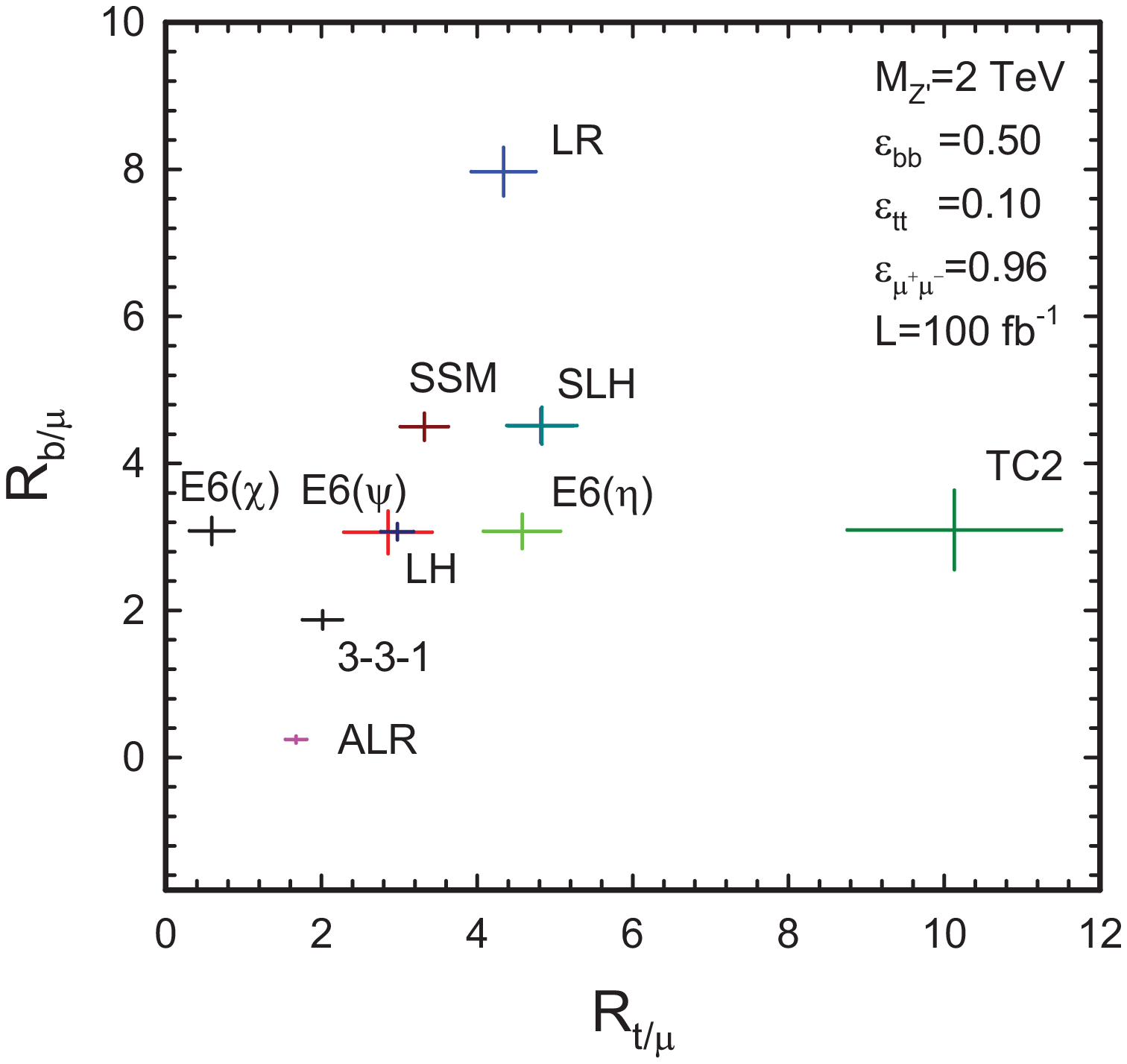} \qquad\qquad\qquad
\includegraphics[width=60mm,clip]{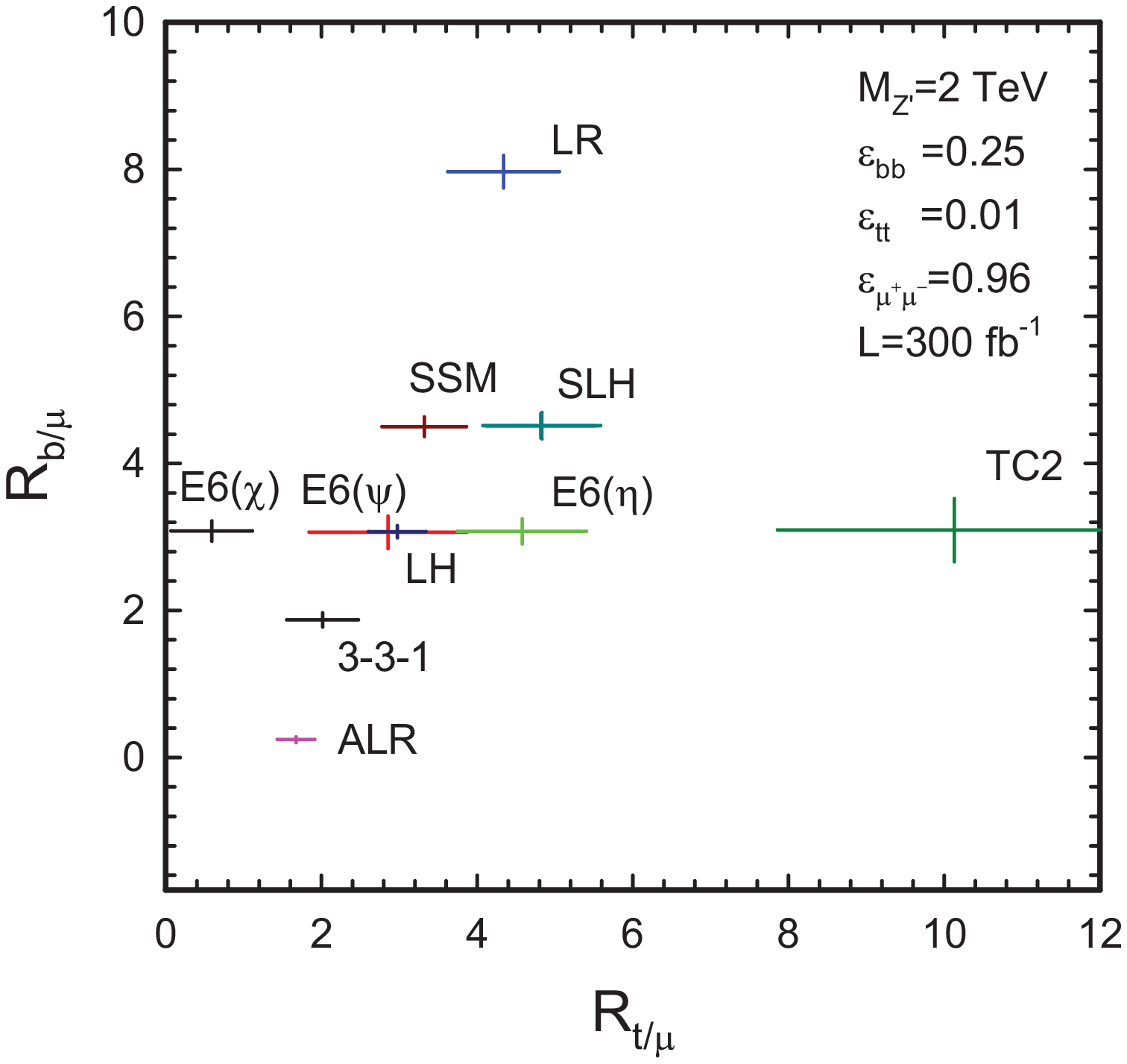}
\caption{Plots of the ratios $R_{b/\mu}$ vs $R_{t/\mu}$ to show the ability to distinguish between 
models. Error bars include statistical uncertainties from signal and background. Kinematic cuts
include $|\eta| < 2.5$ to account for detector tracking, and $p_{T} > 600$~GeV to improve signal
significance over QCD backgrounds.} \label{fig:ratio_plots}
\end{figure*}

In our analysis, we did not consider contributions to uncertainties from the parton distribution
functions and from higher order loop corrections, though the effect of PDF uncertainties can 
be reduced by taking the ratio of cross sections to different final states.
To distinguish between models, we define:
\begin{eqnarray}
R_{b/\mu} & = & {{\sigma(pp\to Z^{\prime} \to b\bar{b})}\over {\sigma(pp\to Z^{\prime} \to \mu^+\mu^-)}} \nonumber \\
& \simeq & { {BR(Z^{\prime}\to b\bar{b})} \over {BR(Z^{\prime}\to \mu^+\mu^-) }} 
= { { 3K_q ( {g_L^b}^2 + {g_R^b}^2 ) }\over { {(g_L^\mu}^2 + {g_R^\mu}^2 ) } } \cr
R_{t/\mu} & = & {{\sigma(pp\to Z^{\prime} \to t\bar{t})}\over {\sigma(pp\to Z^{\prime} \to \mu^+\mu^-)}} \nonumber \\
& \simeq & { {BR(Z^{\prime}\to t\bar{t})} \over {BR(Z^{\prime}\to \mu^+\mu^-) }} 
= { { 3K_q ( {g_L^t}^2 + {g_R^t}^2 ) }\over { {(g_L^\mu}^2 + {g_R^\mu}^2 ) } }
\label{eq-ratio}
\end{eqnarray}
where $K_q$ depends on QCD and electroweak corrections. 
These ratios depend on model dependent couplings that will produce distinctive results based
on the model.

In Fig.~\ref{fig:ratio_plots}, we plot the ratios $R_{b/\mu}$ vs $R_{t/\mu}$~\cite{Godfrey:2008vf} 
for a selection of models for two scenarios representing optimistic and pessimistic views of 
the identification efficiency and background rejection capabilities of the detectors at the LHC.
We assume that a $Z^{\prime}$ has been discovered and a measurement of the mass and width has
been found from the dilepton channel.
The errors shown represent purely statistical uncertainties resulting from subtracting the 
background signal from the total background$+$signal that would be observed.

It should be noted that the unique position of each model in these plots is due to the specific
choice of the mixing parameter value in each case.
In some cases, the mixing parameter is free to vary depending on the nature of the symmetry,
resulting in overlapping regions in the $R_{b/\mu}-R_{t/\mu}$ plane so that one could
not uniquely distinguish the precise model using this measurement alone.
In this case, although it might not be possible to uniquely identify 
the underlying model,  some models could be ruled out and the mixing parameter for others
could be constrained in conjunction with other measurements.

\section{Forward-Backward Asymmetry}

It is possible to take advantage of heavy flavour identification to measure forward-backward
asymmetries ($A_{FB}^{f\bar{f}}$) in $Z^{\prime}$ production.
The forward-backward asymmetry is defined as:
\begin{eqnarray}
A_{FB} & = & {  \displaystyle\frac{ \left[{ \displaystyle\int_o^{y_{max}} - \displaystyle\int_{-y_{max}} ^0 }\right] 
\displaystyle\frac{d\sigma^-}{dy} dy }
{\displaystyle\int_{-y_{max}}^{y_{max}}  \displaystyle\frac{d\sigma^+}{dy} dy} } \cr
& \sim & \displaystyle\left( \frac{{C_L^f}^2 - {C_R^f}^2}{{C_L^f}^2 + {C_R^f}^2}\displaystyle\right)
\left( { { \displaystyle\sum_q G_q^-({C_L^q}^2 - {C_R^q}^2 )} \over { 
\displaystyle\sum_q G_q^+( {C_L^q}^2 + {C_R^q}^2) } }\right),
\label{eq:afb}
\end{eqnarray}
where $C_{L,R}^f$ are the left and right handed couplings of the $Z^{\prime}$ to fermions,
and $G_q^\pm$ are the integrated symmetric and antisymmetric combinations of the parton 
distribution functions.
The differential cross sections, $d\sigma^\pm/dy$, are even and odd contributions to the $Z^{\prime}$ 
rapidity distribution, found via:
\begin{eqnarray}
\displaystyle\frac{d\sigma^\pm}{dy} = \left[\displaystyle\int^{1}_{0} \pm \displaystyle\int^{0}_{-1} 
\right] \displaystyle\frac{d\hat{\sigma}}{dy dz^*} dz^*,
\end{eqnarray}
where the centre-of-mass scattering angle, $z^* = \cos\theta^*$, is measured from the outgoing
fermion relative to the incoming quark.

For $pp$ collisions at the LHC, there is an ambiguity in determining the direction of the quark
where it is impossible to tell on an event-by-event basis whether the $Z^{\prime}$ was boosted 
in the direction of the quark or anti-quark.
Because the momentum distributions are harder for the valence quarks than for the sea 
anti-quarks, this ambiguity can be resolved to a certain extent by assuming that
the $Z^{\prime}$ boost direction is the same as the 
quark direction \cite{Dittmar:1996my}.
For small values of the rapidity of the $Z^{\prime}$, the quark and anti-quark momenta are more evenly
balanced and this procedure is less likely to correctly identify the quark direction.

\begin{figure}[ht]
\centering
\includegraphics[width=60mm,clip]{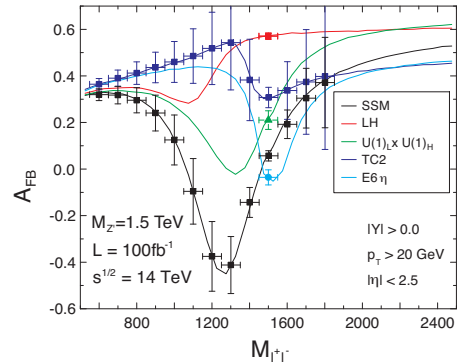}
\caption{Forward-Backward Asymmetry measurements in bins of 100~GeV for a 1.5~TeV $Z^{\prime}$. 
} \label{fig:afb_dist}
\end{figure}

In a recent paper, we suggested that a simpler method of performing the forward-backward
asymmetry measurement is possible by using the direct pseudorapidity measurements of
the final state particles \cite{Diener:2009ee}.
It can be shown that a ``forward" event is one in which $|\eta_f| > |\eta_{\bar{f}}|$ in the lab
frame, and vice-versa for a ``backward" event, assuming that the initial state quark has a
greater momentum than the initial state anti-quark.
This method of finding the $A_{FB}$ has the advantage of being very straightforward
and clean, as it relies only on counting events depending on the pseudorapidity.
No calculation of the centre-of-mass scattering angle or $Z^{\prime}$ rapidity is required.
In Fig.~\ref{fig:afb_dist}, we show the $A_{FB}$ distribution in invariant mass bins of the 
$\mu^+\mu^-$ final state for a $Z^{\prime}$ with mass $M_{Z^{\prime}} = 1.5$~TeV. 

\begin{figure}[t]
\centering
\includegraphics[width=70mm,clip]{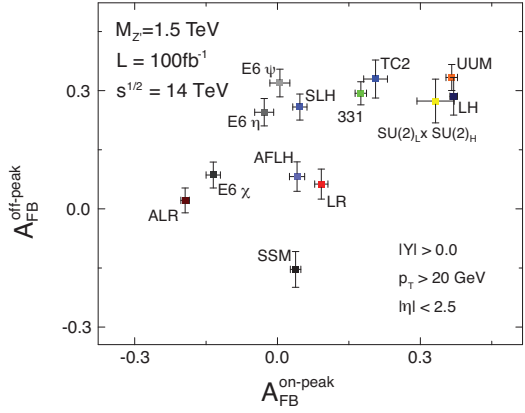}
\caption{Off-peak versus on-peak $A_{FB}$ measurements for a $Z^{\prime}$ with a mass of 1.5~TeV.
The off-peak measurement provides a good companion to the on-peak measurement
to distinguish between models such as the SSM and Anomaly Free Simple Little Higgs model,
for example.} \label{fig:afb_onoff}
\end{figure}

Typically the on-peak $A_{FB}$ has been studied to determine the $Z^{\prime}$ couplings, where
the statistical uncertainty is smallest due to the resonance enhanced cross section.
However, it is clear that information can still be obtained from an off-peak $A_{FB}$ 
measurement~\cite{Petriello:2008zr}.
In Fig.~\ref{fig:afb_onoff} we show the results of an analysis of the on-peak versus off-peak 
$A_{FB}$ for the $\mu^+\mu^-$ final state for a number of models with $M_{Z^{\prime}} = 1.5$~TeV.
In this case, the on-peak measurement includes all events within 
$|M_{\mu^+\mu^-} - M_{Z^{\prime}}| < 3\Gamma_{Z^{\prime}}$ and off-peak includes all events within
the range $2/3 M_{Z^{\prime}} < M_{\mu^+\mu^-} < M_{Z^{\prime}} - 3\Gamma_{Z^{\prime}}$.

\begin{figure}[t]
\centering
\includegraphics[width=70mm,clip]{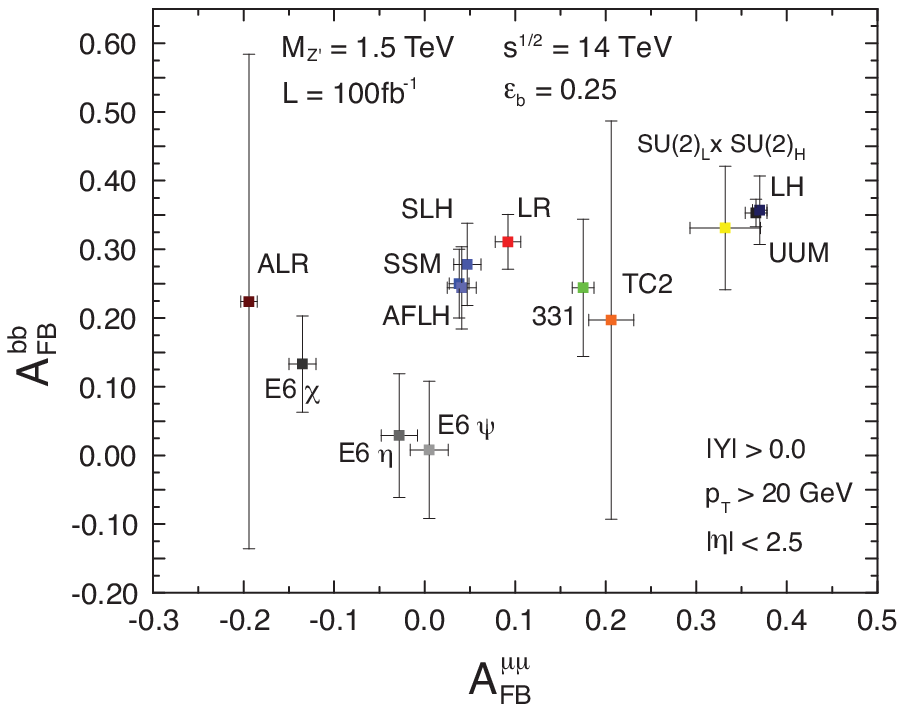} \\
\includegraphics[width=70mm,clip]{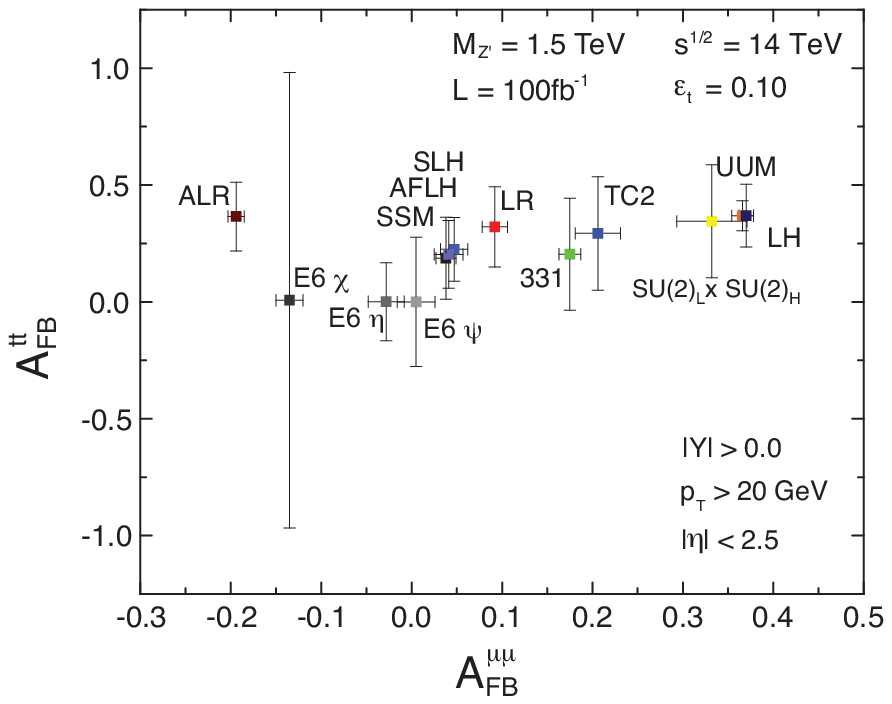}
\caption{$A_{FB}^{b\bar{b}}$ versus $A_{FB}^{\mu^+\mu^-}$ (upper figure)
and 
$A_{FB}^{t\bar{t}}$ versus $A_{FB}^{\mu^+\mu^-}$  (lower figure)
for a $Z^{\prime}$ with a mass of 1.5~TeV. Statistical errors are shown
including contribution from QCD backgrounds.} \label{fig:afb_bbmu}
\end{figure}

In addition to using third generation final states for ratios of cross sections, it may also 
be possible
to measure the forward-backward asymmetry measurement in the $t\bar{t}$ and $b\bar{b}$ channels.
Using pseudorapidity information to make a measurement of an $A_{FB}$ may reduce 
potentially large systematic errors resulting from determining $z^*$ and $y_{Z^{\prime}}$ for hadronic final states.
The large QCD backgrounds will make this measurement difficult, but an $A_{FB}^{q\bar{q}}$ 
measurement would provide additional information about the couplings to $b$ and $t$ quarks
that cannot be obtained from other measurements.
From Eq.~(\ref{eq:afb}), it is clear that the 
term in the second set of brackets does not depend on the final state.
Measurements of $A_{FB}$ to other final states can therefore, potentially, 
give information about the coupling of a $Z'$ to the final state heavy quarks
and will be useful in a global fit of $Z'$ couplings.
Figure~\ref{fig:afb_bbmu} shows $A_{FB}$ for the $b\bar{b}$ 
and  $t\bar{t}$  channels  versus $A_{FB}$ for the the muon channel. 
While large uncertainties are apparent for some models, a reasonable measurement
can still be expected for a number of models including the Left-Right Symmetric Model and 
variants of the Little Higgs that are included.

\section{Summary}

Many models of new physics predict new $s$-channel resonances.
It is possible that a $Z'$ may be discovered early in the LHC program, especially a
$Z^{\prime}$ state decaying to muons or electrons.
Numerous observables have been proposed to distinguish between the different possibilities
of $Z^{\prime}$ models.
In this contribution, we showed that flavour tagging of 3rd generation quarks can be used to
distinguish between models and measure individual quark couplings to a $Z^{\prime}$.
We also described a new method for measuring the forward-backward asymmetry, which
is of particular use in measuring the $A_{FB}$ to third generation quark final states.
Values of $A_{FB}$ were shown for $\mu^+\mu^-$, $t\bar{t}$ and $b\bar{b}$ final states.

\begin{acknowledgments}
This work was supported in part by the Natural Sciences and Engineering Research 
Council of Canada.
\end{acknowledgments}

\end{document}